\documentclass[9pt]{article}

\usepackage{graphicx}
\usepackage{lineno}
\usepackage{chemfig}
\usepackage{indentfirst,csquotes}
\usepackage{lipsum}
\usepackage{authblk}
\usepackage{url}

\title{Beyond spectral resolution in optical sensing: Picometer-level precision with multispectral readout}
\date{\vspace{-3ex}}

\author[1,†,*]{M.S. Cano-Velázquez}
\author[1,†]{S. Buntinx}
\author[1,†]{A.L. Hendriks}
\author[1,†]{A. van Klinken}
\author[1,†,‡]{C. Li}
\author[1]{B.J. Heijnen}
\author[1]{M. Dolci}
\author[1,§]{L. Picelli}
\author[1]{M.S. Abdelkhalik}
\author[2]{P. Sevo}
\author[1,2]{M. Petruzzella}
\author[1,2,3]{F. Pagliano}
\author[1,§]{K.D. Hakkel}
\author[1]{D.M.J. van Elst}
\author[1]{P.J. van Veldhoven}
\author[1,4]{E. Verhagen}
\author[1]{P. Zijlstra}
\author[1]{A. Fiore}

\affil[1]{Department of Applied Physics and Science Education, and Eindhoven Hendrik Casimir Institute, Eindhoven University of Technology, 5600 MB, Eindhoven, The Netherlands}
\affil[2]{MantiSpectra B.V.; Eindhoven, 5612 AE, The Netherlands}
\affil[3]{nanoPHAB B.V.; Eindhoven, 5612 AP, The Netherlands}
\affil[4]{Center for Nanophotonics, AMOLF, Science Park 104, 1098 XG, Amsterdam, The Netherlands}
\affil[$\dag$]{These authors contributed equally to this work.}
\affil[$\ddag$]{Present address: College of Information Science and Electronic Engineering, Zhejiang University; Hangzhou, 310027, China and Research Center for Intelligent Optoelectronic Computing, Zhejiang Lab; Hangzhou, 311121, China}
\affil[$\S$]{Present address: MantiSpectra B.V.; Eindhoven, 5612 AE, The Netherlands}
\affil[*]{m.s.cano.velazquez@tue.nl }

\begin{document}
\maketitle

\begin{abstract}
 Optical sensors offer precision, remote read-out, and immunity to electromagnetic interference but face adoption challenges due to complex and costly readout instrumentation, mostly based on highresolution. This article challenges the notion that high spectral resolution is necessary for high performance optical sensing. We propose co-optimizing the linewidths of sensor and readout to achieve picometer-level precision using low-resolution multispectral detector arrays and incoherent light sources. This approach is validated in temperature sensing, fiber tip refractive index sensing, and biosensing, achieving superior precision than high-resolution spectrometers. This paradigm change in readout will enable optical sensing systems with costs and dimensions comparable to electronic sensors.
\end{abstract}

\section{Introduction}

The distinct feature of optical sensing is the possibility of measuring with high precision at a distance without electrical contact with the object. In some applications, such as spectroscopy and long-distance ranging, the information of interest is directly coded into the optical beam upon reflection from the object. In many other cases, a transducer is needed, which imprints a change into the reflected or transmitted beam, depending on the measurand of interest - this change is then measured by a readout unit. Fiber sensors of physical parameters such as temperature, pressure, acceleration, optical biosensors, and chemical sensors are all based on this combination of transducer and readout. Among the different wave properties that can be used for encoding, the spectrum is by far the most convenient, as it is most tolerant to changes and fluctuations in the transmission channel and to power fluctuations in the light source. Fiber Bragg gratings (FBGs) \cite{Hill1997-vi}, lab-on-fiber sensors \cite{Consales2012-gv}, and many photonic and plasmonic biosensors \cite{Chen2020-gx} use spectral encoding and typically rely on a spectral resonance that shifts due to changes in refractive index within or around the transducer. However, these changes are typically small, in the $10^{-6} - 10^{-2}$ refractive index units (RIU) range, resulting in wavelength shifts from a few picometers (pm) to a few tens of nanometers (nm).

High-resolution spectrometers and tunable lasers have typically been used to measure the resonance wavelength with the required precision, resulting in high complexity and cost. Here, we define resolution as the minimum wavelength difference between two spectral lines that the readout instrument can measure, and imprecision as the root-mean-square error in the measurement of the wavelength of an isolated line. An additional challenge is that single-mode fibers are typically used to carry the signal to the transducer, leading to tight alignment requirements and high packaging costs. As a replacement for the conventional grating-based spectrometer, integrated spectrometers \cite{Sano2003-uy} or spectral-to-spatial mapping and large imaging arrays \cite{Triggs2017-bl, Yesilkoy2019-bb, Pinet2007-cf} have been proposed for sensor readout. Also these approaches involve the measurement of hundreds of spectral/spatial channels and share similar issues in signal handling and system complexity. In turn, the wide availability of high-end spectral instrumentation in research labs has motivated the development of sensors that perform optimally when measured with high resolution (e.g. sensors with minimized linewidth \cite{White2008-mp}). However, as shown below, these design choices do not consider the performance at the system level, nor practical considerations of size and cost. In this paper, we revisit the problem of the spectral readout, considering both the transducer and the readout as parts of a single sensing system. We show that high spectral resolution at the transducer and readout does not improve sensing performance. Instead, a simple, low-resolution multispectral readout approach is introduced, that delivers a wavelength imprecision matching the fundamental Cramér-Rao bound \cite{Chao2016-ei}, a dynamic range suitable for all practical applications, and the possibility of compensating the most common environmental fluctuations. We implement this concept in a small, integrated multispectral detector array and use it, in combination with multimode fibers, to read out three different types of sensors of temperature, refractive index, and biomolecules, respectively. In all cases, we experimentally demonstrate a wavelength imprecision at the picometer level, exceeding the one obtained with a high-resolution lab instrument. This conclusively shows that a simple sensing system, consisting solely of a broadband light source, multimode fibers, an integrated transducer, and a multispectral chip, can provide a sensing performance matching that of high-end instrumentation.

\section{Multispectral readout approach}
\begin{figure*}[ht!]
\centering\includegraphics[width=\linewidth]{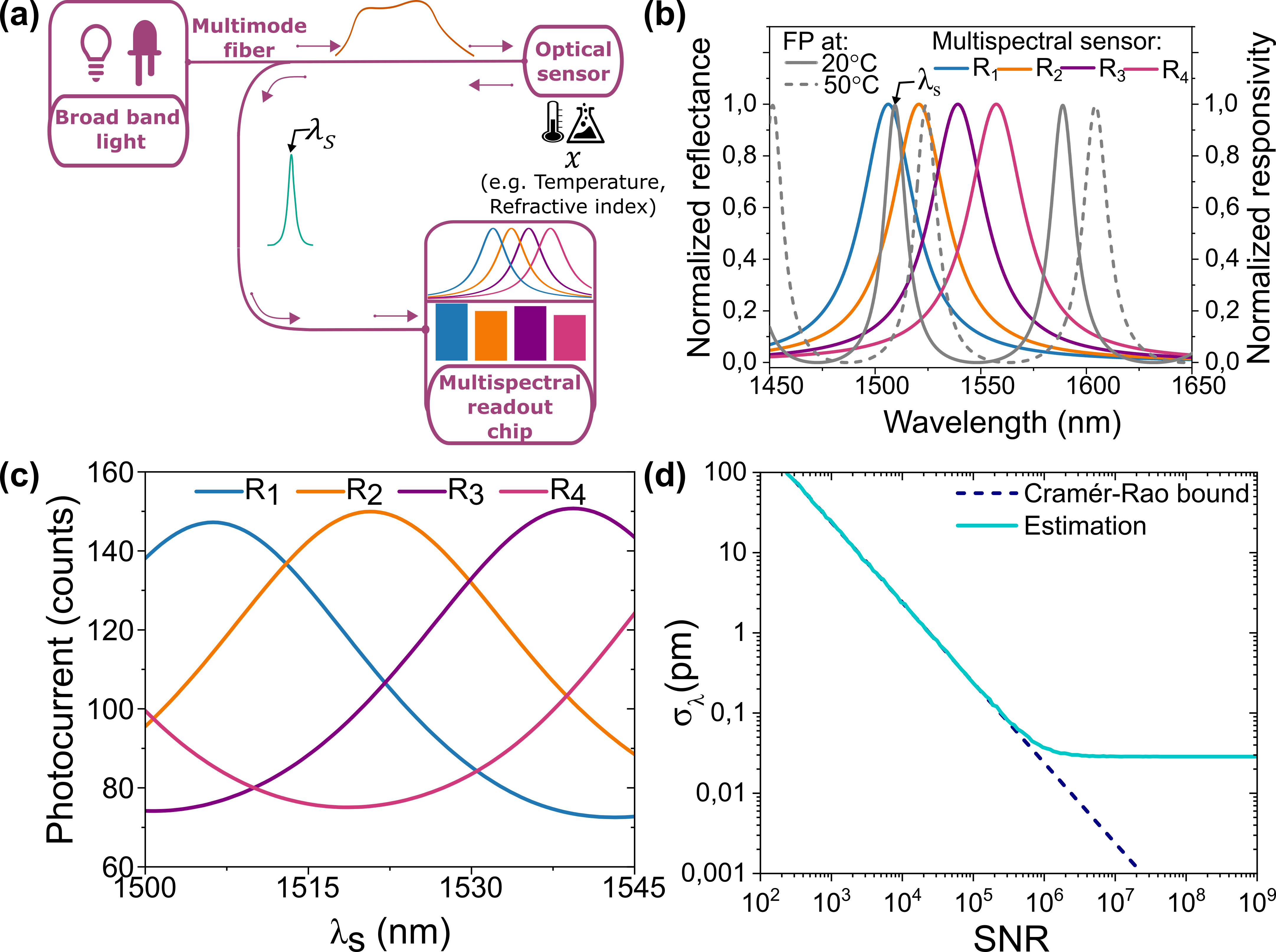}
\caption{Multispectral read-out approach concept. (a) Sketch of the proposed optical system and multispectral readout system. (b) Examples of spectral line shapes of sensor and readout channels. (c) Calculated signal of the four spectral channels of the multispectral readout module vs peak wavelength $\lambda_S$. (d) The standard deviation of wavelength error (estimated $\lambda_S$ - nominal $\lambda_S$) as a function of SNR for 30 calibration points, and comparison with CRLB.}
\label{Fig1:concept}
\end{figure*}

The general problem of spectral readout is illustrated in Fig. \ref{Fig1:concept}(a): The reflection (or transmission) spectrum of a sensor, $R_S(\lambda;x)$, carries information on the measurand $x$. For simplicity, we will assume that the information is coded in the wavelength $\lambda_S$ of a spectral feature (e.g. a peak or dip) which shifts depending on the measurand, $R_S(\lambda;x)=R_S(\lambda;\lambda_S (x))$, but the approach is also applicable to more complex spectral changes. We consider the case of a low-cost sensing system, using an incoherent light source (e.g. a white-light source or a light-emitting diode) and multimode fibers for light delivery, and assume that the source has a nearly uniform power spectral density $P_\lambda (\lambda)$ in the spectral range of interest.  The readout unit consists of a set of $N$ detectors ("spectral channels"), each measuring a part of the spectrum, through a wavelength-dependent responsivity $R_i (\lambda)$ so that their photocurrent is given by $I_i (\lambda_S)=\int_{}^{}R_i (\lambda)R_S(\lambda;\lambda_S) P_\lambda(\lambda)d\lambda$. This describes both the conventional spectrometer-based readout, where the spectral response is defined by the grating, and the multispectral readout proposed in this paper. The sensing goal of determining $\lambda_S$ from the set of photocurrents ${I_i}$ is conceptually analogous to the problem of determining the position of a molecule from a camera image in super-resolution microscopy. However, important differences are that i) the transducer is not a "point emitter", but instead the reflected power scales with its spectral bandwidth; and ii) we can in principle engineer the spectra of both sensor and readout unit at will. Drawing from general information theory results, the variance of any unbiased estimator of $\lambda_S$  (i.e. the fundamental limit to the wavelength imprecision) is limited by the Cramér-Rao lower bound (CRLB), related to the Fisher information matrix \cite{Chao2016-ei}. Assuming that the noise currents in the $N$ detectors are uncorrelated and Gaussian, with equal standard deviation $s_I$ (case of dominant thermal noise, see Supplement 1 for the case of shot noise), the CRLB is given by:

\begin{equation}
\mathrm{\sum}_{\lambda_S}^{CR}=\frac{\sigma_I}{\sqrt{\sum_{i=1}^{N}\left(\frac{\partial \overline{I_i}}{\partial \lambda_S}\right)^{2}}}
\label{eq1:concept}
\end{equation}

The derivatives $\frac{\partial \overline{I}_i}{\partial \lambda_S}=\int_{}^{} R_i (\lambda)\frac{\partial R_S}{\partial \lambda_S} P_\lambda(\lambda)d\lambda$ quantify the contribution of each detector to the Fisher information and thereby the sensitivity to spectral changes. It is straightforward to conclude from Eq. \ref{eq1:concept} that \emph{the imprecision is minimized by co-optimizing the sensor and readout} - for example, choosing spectral lines with equal linewidth, which maximizes the sensitivities $\frac{\partial \overline{I}_i}{\partial \lambda_S}$ (see Supplement 1). An additional observation is that, for this optimal choice, the imprecision does not depend on the linewidth, but it is fundamentally limited by the source power spectral density and the detector's noise equivalent power $P_{min}$, as $\sigma \mathrm{}_{\lambda_S}^{CR}\sim \frac{P_{min}}{P_\lambda}$. This is at odds with the commonly accepted paradigm that high resolution in sensor and readout is needed for a precise wavelength measurement and with the corresponding design rules for resonant sensors \cite{White2008-mp, Conteduca2022-sz}. In particular, the common practice of using a spectrometer with a resolution much smaller than the sensor's linewidth results in a degraded wavelength imprecision. Here we propose the use of a few spectral channels (i.e. detectors integrated with filters within a single multispectral chip), solving the dynamic range and practical limitations of single-channel \cite{Paulsen2017-qd} and two-channels \cite{Davis1994-ce} readout approaches based on discrete filters and detectors, and the complexity of multiple-source based readout \cite{Daaboul2011-cs}. 

\vspace{0.35cm}
We specifically consider four spectral channels, a sensor with a spectral linewidth in the few tens of nm range, and a light-emitting diode or a halogen lamp as a light source. This allows the use of a large-core fiber with a high numerical aperture and uncritical spectral alignment between the sensor and readout. The key idea behind our approach is that this simple configuration, involving only low-cost components and inexpensive packaging, can provide a wavelength imprecision matching the one of high-end lab instrumentation. We first numerically evaluate the expected performance when the sensor consists of a 10 mm-thick Fabry-Perot (FP) cavity between two thin metal mirrors, and the readout channels display Lorentzian lines with a linewidth of 32 nm (Fig. \ref{Fig1:concept}(b)). This choice of linewidth allows placing four overlapping spectral channels with the emission spectrum of a near-infrared light-emitting diode (full-width half-maximum $ \sim$ 100 nm). The goal is to retrieve the peak wavelength $\lambda_S$ of one of the FP peaks, which is an assumed function of a measurand, e.g. temperature. The detector photocurrents are calculated (Fig. \ref{Fig1:concept}(c)) for different temperatures by integrating the product of the FP reflectance spectrum and the detector's responsivities, while  adding Gaussian noise. Once the map is known, a model to retrieve the peak wavelength from the intensity read by each spectral channel is developed, through an iterative minimization algorithm (see Supplement 1). 

Repeating the prediction for different noise realizations provides the standard deviation of the predicted temperature for each signal-to-noise ratio (SNR), which is calculated as the average of the ratio of the photocounts of each channel divided by its standard deviation (Fig. \ref{Fig1:concept}(d): $\sigma_\lambda$ vs SNR). The close match between the standard deviation of the prediction and the CRLB, $\sigma \mathrm{}_{\lambda_S}^{CR}$, shows that the regression model works optimally. A wavelength imprecision $\sigma_{\lambda_S}<$ 1 pm, many orders of magnitude smaller than the linewidth ($\sim$ 12 nm) of the sensor, can be obtained with an $SNR>2.4 \times 10^4$. Besides resulting in a larger dynamic range, the four channels allow correcting for the influence of other environmental parameters. For example, the employed regression model provides a wavelength estimation that is independent of the total power from the source. Additionally, the sensor can be designed in such a way that different physical parameters produce different changes in the photocurrents so that they can be simultaneously retrieved by a multiparameter estimation algorithm. We note that the noise floor in the estimation is entirely controlled by the number of calibration points (30 in the simulation of Fig. \ref{Fig1:concept}(d)), and can in principle be reduced at will.

\section{Integrated multispectral readout chip}

\begin{figure*}[ht!]
\centering\includegraphics[width=\linewidth]{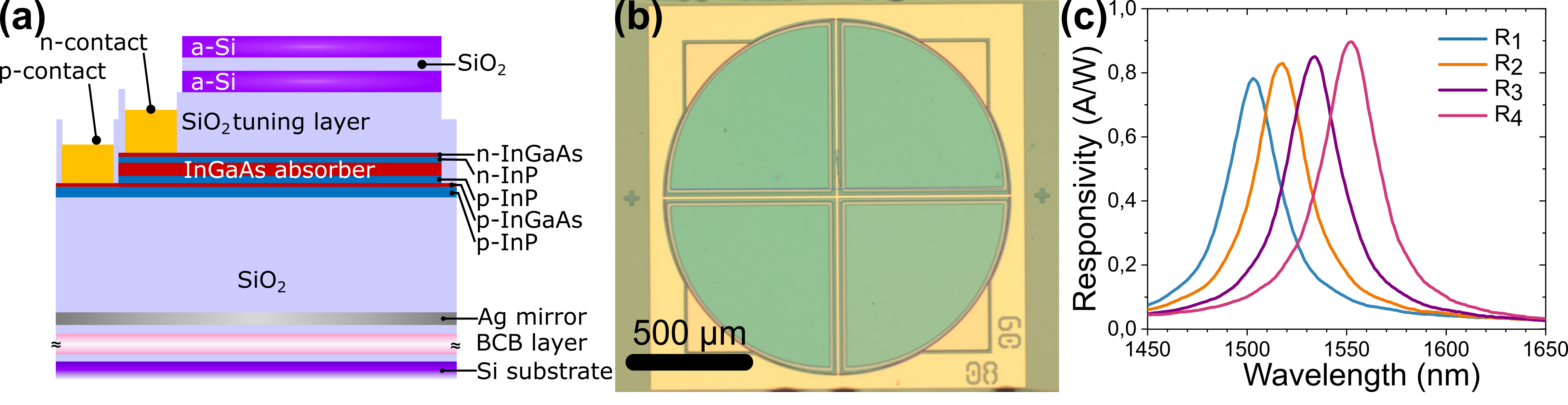}
\caption{Integrated multispectral chip. (a) The device structure of a single resonant-cavity-enhanced photodetector. (b) Optical image of fabricated multispectral readout chip, consisting of four different photodetectors in a pie geometry. (c) Measured responsivity curves of the four photodetectors in the spectral range of interest.}
\label{Fig2:readout}
\end{figure*}

The integrated multispectral readout is implemented with an array of four resonant-cavity-enhanced photodetectors (RCE-PD), based on \chemfig{InP}-on-silicon photodiode technology \cite{Hakkel2022-qr,Van_Klinken2023-tc}. The structure is shown in Fig. \ref{Fig2:readout}(a) and consists of an \chemfig{InP}/\chemfig{InGaAs} p-i-n diode inside of a Fabry-Pérot cavity, formed by a bottom \chemfig{Ag} mirror and a top Bragg mirror composed of 1.5 pairs of amorphous \chemfig{Si} and \chemfig{SiO_2} layers (see Supplement 1). The thickness of a \chemfig{SiO_2} tuning layer between the mirrors is varied to define four responsivity peaks in the spectral range of interest (1450 – 1600 nm), spaced by about 20 nm with respect to each other. The four photodetectors are arranged in a pie geometry and the optically active area has a diameter of 1.3 mm, as seen in Fig. \ref{Fig2:readout}(b). The four detectors have a common p-contact and separate n-contacts. The responsivity curves (Fig. \ref{Fig2:readout}(c)) feature peak responsivities from 0.78 A/W to 0.9 A/W, with linewidths between 32.7 nm to 34.2 nm (full-width half-maximum, FWHM). This very compact multispectral readout chip provides a set of spectral responses close to optimal for the readout of the different sensors described below. A readout board equipped with a 24-bit Analogue to Digital Converter (ADC) was used to measure the photocurrents from the multispectral chip.

\section{Sensing applications}
\subsection{Temperature sensing}

\begin{figure*}[h!]
\centering
\includegraphics[width=\linewidth]{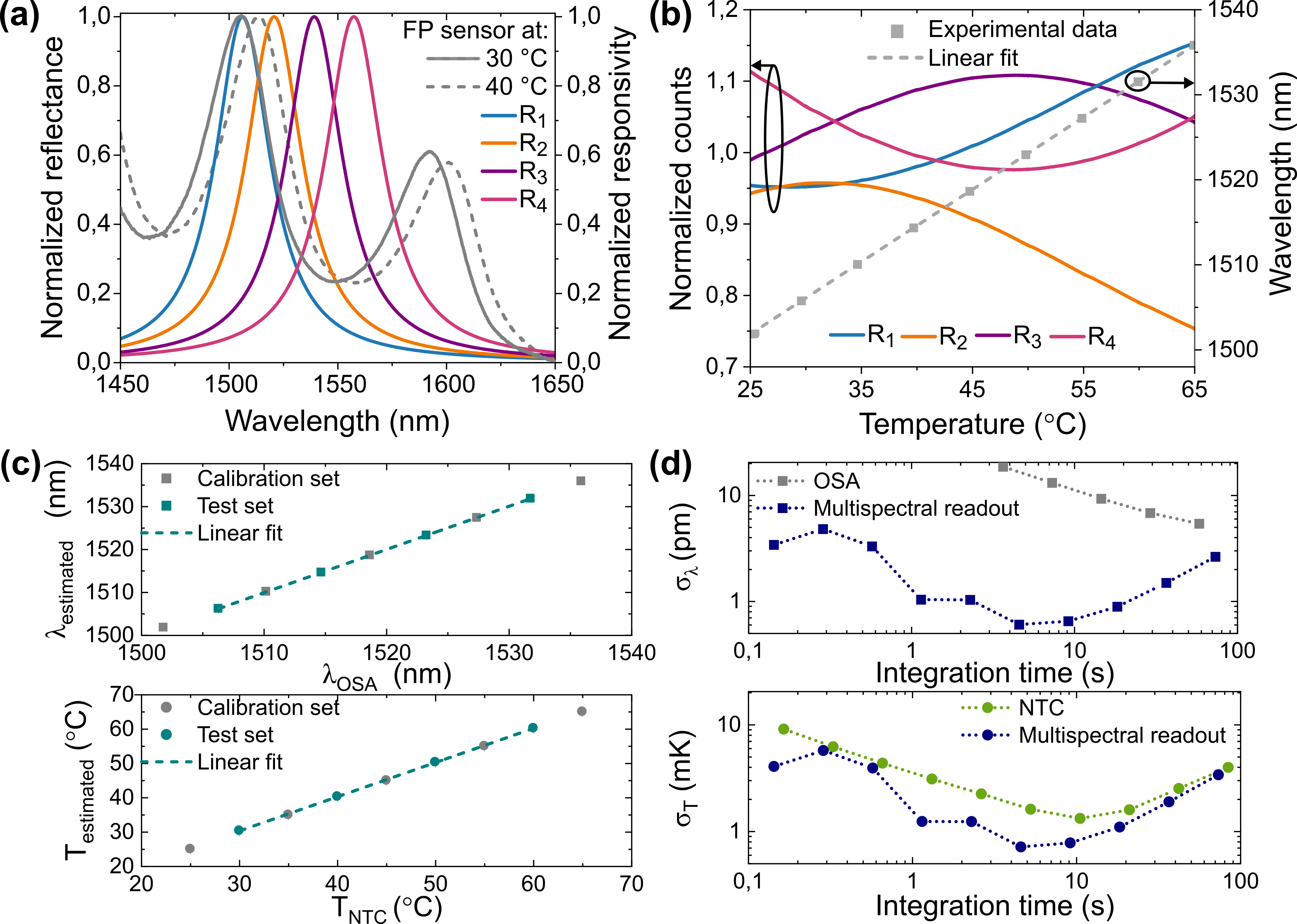}
\caption{Temperature sensing results. (a) Reflected spectrum from the FP sensor, together with the normalized responses of the multispectral photodetectors. (b) Variation of photocurrents (left axis) and OSA-fitted wavelength (right axis) as a function of temperature. (c) Wavelength predicted with multispectral readout vs OSA-fitted wavelength and temperature predicted (directly) with multispectral readout vs independently measured temperature. (d) Allan deviation of wavelength and temperature for multispectral readout, OSA, and reference temperature sensor.}
\label{Fig3:temperature}
\end{figure*}

For temperature sensing, an extrinsic Fabry-Pérot (FP) cavity is used as a sensing element. The cavity consists of a polydimethylsiloxane (PDMS) membrane positioned between two gold layers (thicknesses 10 $\mu$m and 5 nm, respectively), which serve as mirrors. PDMS is chosen due to its high thermal expansion coefficient\cite{Wolf2018-eh}. To couple light to and from the FP cavity, multimode fibers and two couplers with identical characteristics (400 $\mu$m core diameter and 0.39 NA) are utilized. The sensor response is then directed to an Optical Spectral Analyzer (OSA) and the multispectral sensor. Figure \ref{Fig3:temperature}(a) illustrates the normalized reflectance of the sensor as measured by the OSA at two different temperatures, along with the normalized multispectral read-out responses. The interference pattern from the FP cavity can be observed to red-shift at a rate $\frac{\partial \lambda_S}{\partial T}=0.85\frac{nm}{^\circ C}$ (Fig. \ref{Fig3:temperature}(b), right axis), leading to a variation in the photocounts detected by each pixel (Fig. \ref{Fig3:temperature}(b), left axis).

To calibrate and test the estimation method, the OSA spectra, the temperature readings from a reference Negative Temperature Coefficient (NTC) thermistor, and the photocounts of each pixel were measured at various stable temperatures (see Supplement 1). Five temperature points were used as calibration data, while four different points were used for the test set. 

Two estimation models are built, one to predict the wavelength, using the peak wavelength fitted from the OSA spectra as a reference, and the other to predict the temperature directly from reference temperatures from the thermistor (Fig. \ref{Fig3:temperature}(c)). In both cases, an excellent correlation (calculated for the test data) is found, with $R^2$ values of 0.9999 for both cases. This highlights the performance and accuracy of the multispectral readout estimation across different datasets, reinforcing its reliability and robustness. Furthermore, an Allan deviation analysis \cite{Jerath2018-et} was conducted on a series of data points taken in temperature-stable conditions (Fig. \ref{Fig3:temperature}(d)). As expected, for short integration times $t$ the imprecision in wavelength and temperature scales approximately as $\frac{1}{\sqrt{t}}$, as the noise is dominated by the readout noise, while they become limited by long-term drifts at longer times. However, the wavelength imprecision provided by the multispectral readout is over one order of magnitude lower than the one obtained by fitting the OSA spectra for a comparable integration time. This is fully in line with the general expectation from the Cramér-Rao bound (Eq. \ref{eq1:concept}): Spreading the incoming light over more spectral channels results in a lower sensitivity per channel, and thereby higher imprecision. The multispectral readout achieves an ultralow wavelength imprecision of 0.6 $pm$ with an integration time of 4.5 $s$ - this level of resolution, typical of sensing systems based on single-mode fibers, narrow-linewidth FBG sensors, and the corresponding interrogation \cite{hyperion, Optics11}, is obtained here with a simple and inexpensive hardware with minimal packaging requirements. The temperature imprecision, obtained directly from the temperature estimation algorithm, has a minimum of 0.7 mK at 4.5 s integration time, well below the imprecision of the reference thermistor and matching the best values obtained with FBGs and FBG interrogators \cite{De_Landro2020-js,fbgluna}. These values, obtained with an SNR of $\sim 1.5 \times 10^5$, match well the CRLB value calculated from the experimental sensitivities (see Supplement 1). The use of a light-emitting diode as a light source was also tested, and shown to provide a similar wavelength and temperature imprecision of 1.2 pm and 1.5 mK, respectively, due to a lower $SNR= 4.9 \times 10^4$ (see Supplement 1).

\subsection{Fiber-tip refractive index sensing}

\begin{figure*}[h!]
\centering
\includegraphics[width=\linewidth]{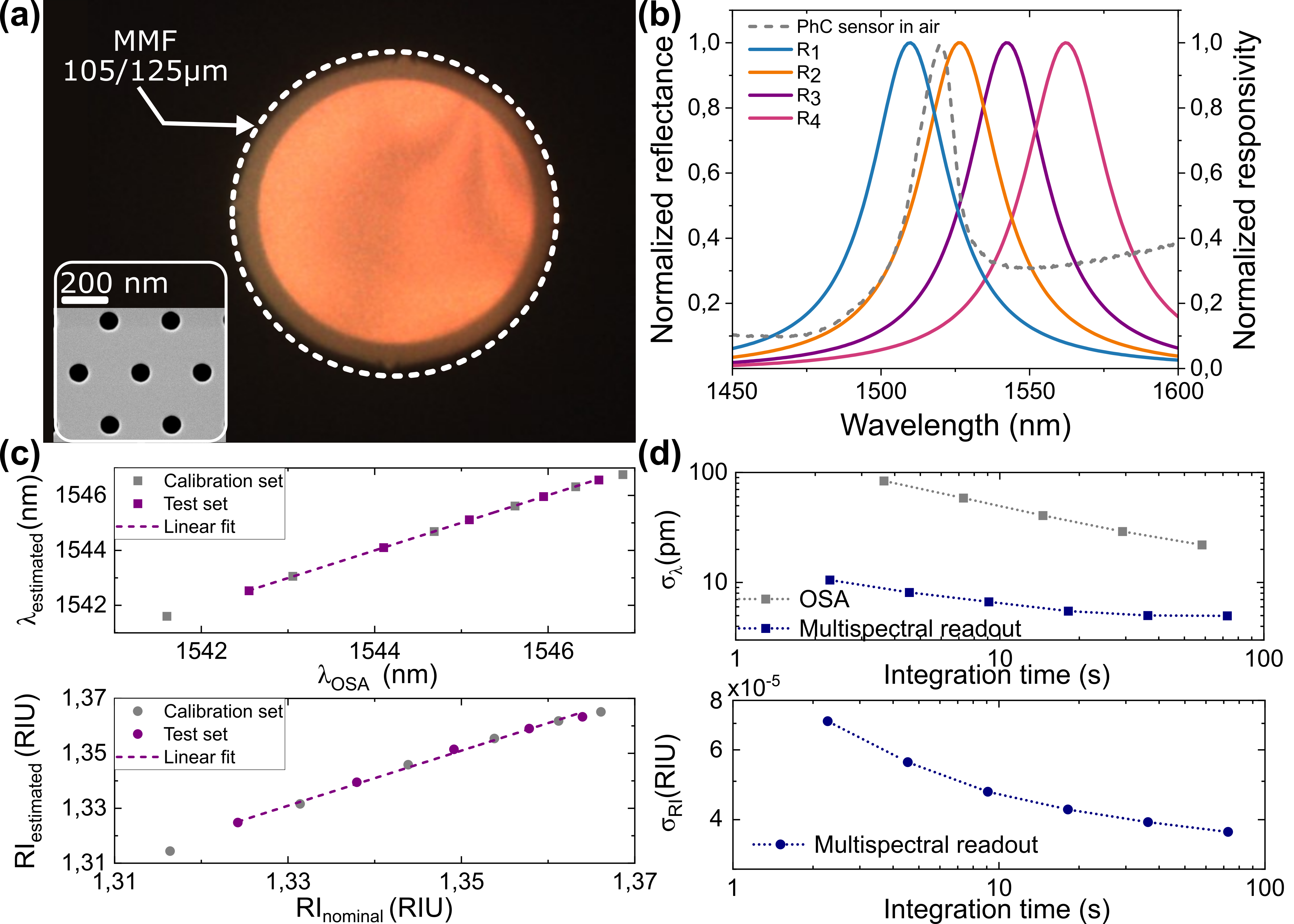}
\caption{Fiber-tip refractive index sensing results. (a) Microscope image of a PhC on fiber tip and SEM image of the hexagonal PhC (inset). (b) Reflected spectrum from the fiber-tip PhC sensor in air, together with the normalized responses of the multispectral photodetectors. (c) Wavelength predicted with multispectral readout vs OSA-fitted wavelength and refractive index predicted (directly) with multispectral readout vs nominal value. (d) Allan deviation of predicted wavelength and refractive index.}
\label{Fig4: RIsensing}
\end{figure*}

As a second application example, we show the use of multispectral readout to interrogate fiber-tip photonic crystal sensors of refractive index, which can be translated to the concentration of an analyte in a solution. Due to their minimal footprint, these sensors can be used for the in-line monitoring of chemical and biochemical processes in mini- and micro-reactors\cite{Li2020-dh}. Unlike the temperature sensor, the fiber-tip sensor has the sensing element directly onto the end-face of the fiber and consists of a 2D photonic crystal slab (PhC). Fiber-tip PhC sensors have been previously demonstrated using single-mode fibers and commercial interrogators or spectrometers\cite{Park2011-lr, Picelli2020-ze,Cano-Velazquez2023-up}. Here, we define a 120 $\mu$m-diameter PhC (Fig. \ref{Fig4: RIsensing}(a)) with a hexagonal lattice and transfer it on top of an MM fiber with a core size of 105 $\mu$m and a numerical aperture of 0.22. The employed PhC lattice design features polarization-degenerate modes with low angular dispersion\cite{Conteduca2021-ic}, which makes it well-suited for use with multimode fibers. The PhC is defined on a 250 nm-thick InP membrane using wafer-scale nanofabrication methods, and afterwards, the PhC is mechanically transferred to the fiber tip using the process described in Ref. \cite{Picelli2020-ze} (see Supplement 1). The reflectance from the fiber-tip PhC sensor (Fig. \ref{Fig4: RIsensing}(b)) shows a clear peak centered at $\lambda_0$= 1519.6 nm, with a FWHM = 18.4 nm.

To evaluate the refractive index sensing performance, experiments were conducted using mixtures of deionized water (DI) and isopropanol (IPA) with varying concentrations. While maintaining a constant temperature, the sensor was immersed in eleven different solutions, from which six concentrations were used as calibration data and five as test set. The estimation model consistently and accurately estimates the peak wavelength across the entire refractive index range (Fig. \ref{Fig4: RIsensing}(c)), with an $R^2$=0.999. For the direct estimation of the refractive index on the test data, an $R^2$=0.997 was obtained. The Allan deviation analysis (Fig. \ref{Fig4: RIsensing}(d)) displays a minimum wavelength imprecision of 5 pm at 72 s integration time. This is higher than for the FP temperature sensor, due to the lower optical power coupled to the 105 $\mu$m-diameter fiber-core. The wavelength imprecision obtained with the multispectral readout is again much lower than the one obtained by fitting the OSA spectra. The refractive index imprecision, as derived directly from the RI estimation algorithm, has a minimum value of $\sigma_{RIU}=3.7 \times 10^{-5}$ RIU, for an integration time of 72 s, and the corresponding limit of detection (LoD) is estimated as LoD = 3$\sigma_{RIU}$ = $1.1 \times 10^{-4}$ RIU.

\subsection{Biosensing}
The third sensing application we analyze is the specific detection of biomarkers in a solution. Immunoglobulin G (IgG) is used as a capture probe to detect its anti-IgG in phosphate-buffered saline (PBS). The biosensor is based on a 1 mm-diameter 2D photonic crystal (PhC) structure, consisting of a hexagonal pattern of holes in a SiN layer on a glass substrate, as shown in the inset of Fig. \ref{Fig5: biosensing}(b) (see Supplement 1). Such PhCs have already been used to demonstrate specific detection of biomarkers and viruses \cite{Triggs2017-bl, Paulsen2017-qd, Inan2017-qk}. Here we show that a multispectral chip enables high-precision readout in a very simple optical system. In this case, light from the fiber is focused onto the PhC within a fluidic cell using two lenses. As seen from the reflection spectrum of the PhC measured with an OSA (Fig. \ref{Fig5: biosensing}(a)), the additional loss, related to free-space coupling, makes the reflectance modulation from the PhC comparable to the background reflectance from the fiber beamsplitter. Also, the PhC structure has a complex reflection spectrum consisting of several peaks, which originate from different optical modes. This configuration is therefore a more challenging test for the multispectral readout. 

To calibrate the multispectral readout to the response of the biosensor, the bulk refractive index is varied by exposing the PhC structure to ethylene-glycol solutions of different concentrations before the start of the biosensing experiment. The wavelength of the largest peak at around 1528 nm is used as a reference for the calibration. In Figure \ref{Fig5: biosensing}(c), the time trace of the wavelength predicted by the multispectral readout during the calibration (0 to 55 min) and biosensing experiment (after 55 min) is plotted together with the actual peak wavelength determined from the simultaneously measured spectrum. 

After calibration, the bioassay includes sequential steps starting with the physisorption of the capture probe (IgG at 300 nM), during which a rapid red-shift (0.6 nm) is observed indicating the saturation of the PhC surface by IgG. This step is followed by the adsorption of casein as a blocking agent to suppress nonspecific interactions. Finally, the biomarker (anti-IgG) is injected at a concentration of 100 nM and causes a red-shift of 0.5 nm. Subsequent rinsing with PBS does not result in a detectable blue-shift, indicating a strong interaction between the capture probe and the biomarker. Control experiments shown in Supplement 1 confirm the specificity of the signals. The resonance shifts measured with the multispectral readout show a good match with the prediction of the actual peak wavelength, despite the complex spectral shape. The minor deviations observed are attributed to the calibration based on the bulk refractive index sensing data, which cannot fully describe the surface binding process (as the different optical modes have different surface/bulk sensitivities).

From the time trace of the predicted wavelength in nominally stable conditions, the Allan deviation is determined, as shown in Fig. \ref{Fig5: biosensing}(b), showing a wavelength imprecision of 5 pm for integration times of a few seconds, which is remarkable given the complexity of the PhC reflectance spectrum and the presence of significant background. Given the measured bulk refractive index sensitivity of 235 nm/RIU, this wavelength imprecision translates into a LoD = $6.4 \times 10^{-5}$ RIU, lower than the one obtained with intensity-based readout \cite{Paulsen2017-qd} and comparable to values obtained with spectral-spatial readout \cite{Kenaan2020-wd}, which requires a camera and a free-space imaging system. Note that the properties of the assay are determined by the affinity and rate constants of the biomolecular interaction. Improvements in kinetic response and sensitivity, therefore, require screening of both antibodies and the sensor’s surface chemistry \cite{Welch2017-rz}. Combining these improvements with a better fiber-PhC coupling, leading to sub-pm wavelength imprecision, will enable pM-level biosensing assays with hardware complying with the footprint and cost requirements of point-of-care diagnostics.

\begin{figure*}[h!]
\centering
\includegraphics[width=0.98\linewidth]{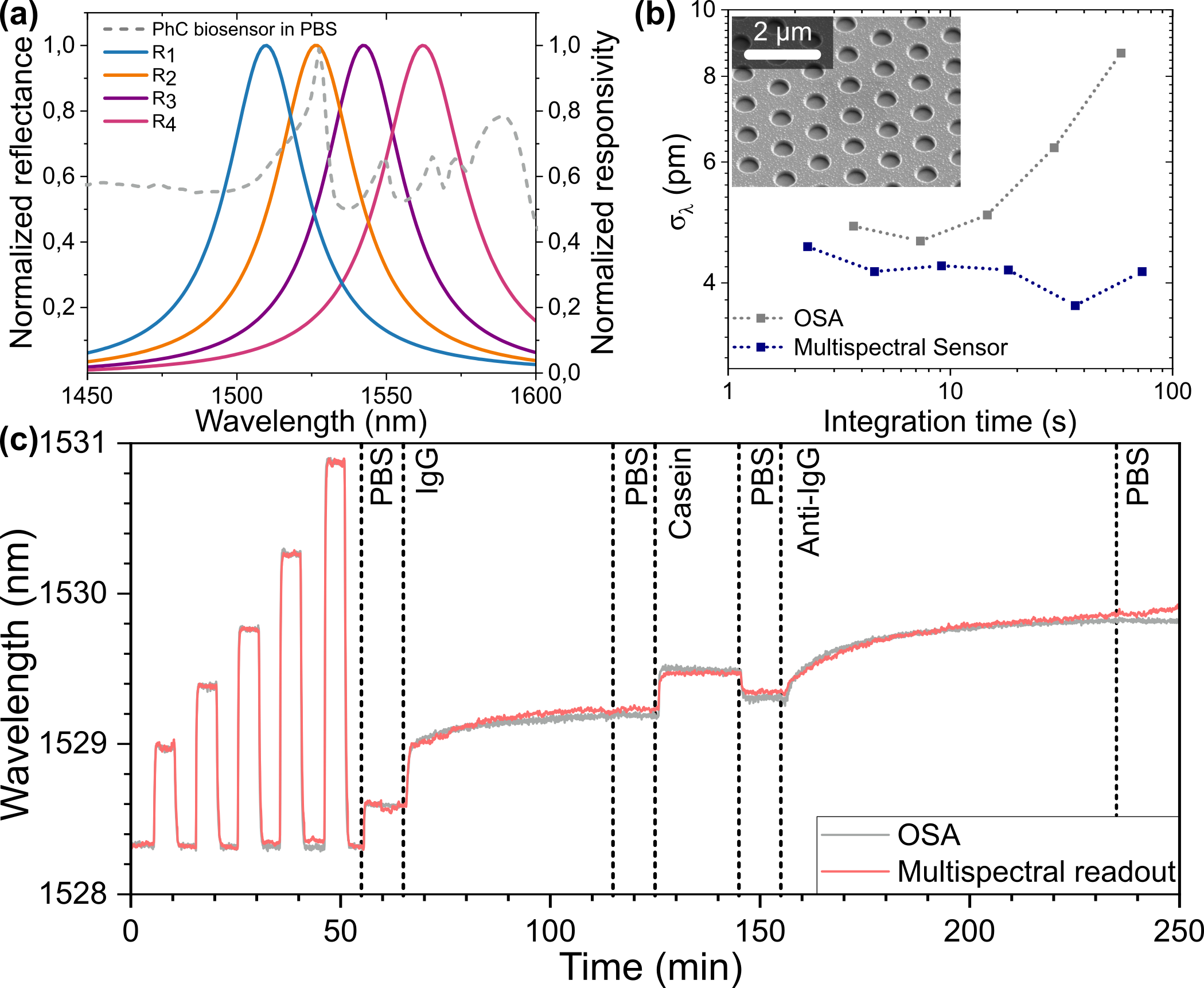}
\caption{Biosensing experiment results. (a) Reflected spectrum from the photonic crystal biosensor in PBS, together with the normalized responses of the photodetectors. (b) Allan deviation of wavelength prediction for multispectral readout (inset: SEM image of photonic crystal biosensor). (c) Time trace of the resonance wavelength determined from the OSA and multispectral readout during a biosensing experiment, including an RI sensing series for calibration (0 to 55 minutes).}
\label{Fig5: biosensing}
\end{figure*}

\section{Conclusions}
Starting from an analysis of the fundamental limits, we have established a general framework for comparing the performance of optical sensing systems that transduce the parameter of interest into a spectral change. This approach considers the complete source/transducer/readout system and provides simple figures of merit (the sensitivities in Eq.\ref{eq1:concept}) that apply to different readout hardware, without any assumption on the data analysis. It shows that the transducer and readout must be co-optimized. Moreover, contrary to the common understanding, in the situation of a broad light source which is most relevant for low-cost sensing applications, minimizing the transducer and readout linewidth does not lead to improvements in the sensing performance. Motivated by these findings, we have developed an integrated multispectral readout chip, which, combined with broad light sources and multimode fibers, enables wavelength measurements with a picometer-level precision. This approach was tested in three different sensing problems, with various transducer structures, delivering a sensing performance comparable to the one obtained using high-end fiber-optic instrumentation. This opens the way to affordable, optical sensing systems based exclusively on integrated components, which can be packaged inexpensively due to the use of large-core fibers. Further optimization of the readout electronics and the design and assembly of sensors may lead to wavelength imprecisions at the 100 fm level and a subsequent order-of-magnitude improvement in the detection limits.

\vspace{0.5cm}

\textbf{Funding} Nederlandse Organisatie voor Wetenschappelijk Onderzoek (NWO) Toegepaste en Technische Wetenschappen (TTW) project n. 16670, project n. 18477 and project n. 17626. NWO Nationale Wetenschapsagenda (NWA) Kleine Projecten NWA.1418.22.022. NWO Zwaartekracht Research Center for Integrated Nanophotonics. European Union’s Horizon 2020 grant n. 864772. 
\vspace{0.2cm}

\textbf{Acknowledgments} We are grateful to Prof. Thomas Krauss and Prof. Sjoerd Stallinga for the valuable discussions. This research was partially funded by the Nederlandse Organisatie voor Wetenschappelijk Onderzoek (NWO) Toegepaste en Technische Wetenschappen (TTW)  project n. 16670 (A.v.K., C.L., P.J.v.V.), the NWO TTW project n. 18477 (A.L.H., M.S.C, M.D.), the NWO TTW project n. 17626 (D.M.J.v.E.), the NWO Nationale Wetenschapsagenda (NWA) Kleine Projecten NWA.1418.22.022 (M.S.C.), the NWO Zwaartekracht Research Center for Integrated Nanophotonics (L.P.),  the European Research Council (ERC) under the European Union’s Horizon 2020 grant n. 864772 (P.Z.) and the Research programme of the Netherlands Organisation for Scientific Research (NWO) (E.V.). We also express our gratitude to the University Fund Eindhoven who contributed to the research with philanthropic funding from alumni.
\vspace{0.2cm}

\textbf{Disclosures} A. Fiore, M. Petruzzella, and F. Pagliano are cofounders and shareholders of MantiSpectra. A. Fiore and M. Petruzzella are inventors on a patent titled "A spectral sensing system" (Patent Number: WO2022/265497).
\vspace{0.2cm}

\textbf{Data Availability Statement} Data underlying the results presented in this paper are not publicly available at this time but may be obtained from the authors upon reasonable request.

\vspace{0.2cm}
\textbf{Supplemental document} See Supplement material for supporting content.

\bibliographystyle{unsrt}
\bibliography{BeyondSP}



\end{document}


\maketitle

\section{Fabrication processes}

\subsection{Multispectral sensor}
The p-i-n structure is firstly grown on a 2" indium phosphide wafer using metalorganic vapor-phase epitaxy (MOVPE) and consists of an n-InGaAs contact layer (30 nm, doped with Si at $1\times 10^{19}$ cm$^{-3}$), an n-InP spacer layer (50 nm, doped with Si at $5\times 10^{18}$ cm$^{-3}$), an InGaAs absorber layer (100 nm, non-intentionally doped), a p-InP spacer layer (50 nm, doped with Zn at $1\times 10^{18}$ cm$^{-3}$), a p-InGaAs contact layer (30 nm, doped with Zn at $1\times 10^{19}$ cm$^{-3}$) and a p-InP support layer (80 nm, doped with Zn at $1\times 10^{18}$ cm$^{-3}$). After cleaving and cleaning the wafer, a 908 nm-thick optical spacer SiO$_2$ layer is deposited by plasma-enhanced chemical vapor deposition (PECVD) to achieve resonant modes in the 1450-1600 nm range. A bottom mirror consisting of a 2nm/100nm/2nm Ge/Ag/Ge layer is evaporated. The InP wafer is then flipped and bonded on a silicon substrate using benzocyclobutene (BCB). After that, the InP substrate and the InGaAs and InP capping layers are removed by wet etching, so that the p-i-n membrane structure is exposed. Through optical lithography, a mesa is defined and wet-etched, forming the optically active area of each single detector. Using PECVD, a 492 nm-thick SiO$_2$ layer is deposited. In two consecutive lithography steps, two different photodetectors are selected and in the optically active area, the thickness of the SiO$_2$ layer is reduced by 20 nm or 40 nm, respectively, by wet etching with buffered HF (BHF). In this way, four photodetectors with four different, equally-spaced optical path lengths are created, leading to different wavelengths of the resonant modes. P and N contacts are then defined by two additional steps, where the SiO$_2$ layer is removed by wet etching and the metal contacts are created using a lift-off process. In the final lithography step, the top Bragg mirror is deposited, consisting of 1.5 pairs of amorphous silicon (165 nm) and SiO$_2$ (100 nm). The fabricated diodes show a low dark current of 3 nA at a reverse bias of -0.5V. The fabricated multispectral readout chips are diced and wire-bonded onto chip carriers, which are compatible with a custom-made electronic readout board.

\subsection{Fabry-Pérot temperature sensor}
The Fabry-Pérot cavity (FP) is fabricated on top of a silicon substrate. The first mirror consists of  2 nm/20 nm/2 nm Ti/Au/Cr layers which are evaporated using electron beam evaporation. This is followed by the spin coating of a 10 $\mu$m-thick film of polydimethylsiloxane (PDMS) and crosslinker (10:1). Then, the polymer is cured for 1 h at 200 $^\circ$C. Afterwards, the second mirror is evaporated, consisting of 2 nm/20 nm Ti/Au. Finally, the wafer is diced into 1$\times$1 mm$^2$ squares. 

\subsection{Photonic crystal fiber-tip refractive index sensor}
The photonic crystal (PhC) and surrounding support structures are fabricated using standard wafer-scale growth, lithography, and etching techniques. A 250 nm thick InP membrane which is separated from an InP [100] substrate by a 300 nm-thick InGaAs sacrificial layer were grown by MOVPE. Both layers are lattice matched to avoid strain in the final etching steps. Using a 200 nm-thick hard mask of SiN deposited by PECVD and a spin-coated ZEP520A resist the PhC and supporting structures are patterned with electron beam lithography (EBL). Then, the patterns are transferred into the hard mask and InP membrane by reactive ion etching (RIE) and inductively-coupled plasma RIE (ICP-RIE) respectively. This is followed by the removal of the SiN by BHF and the deposition of SiN on both sides of the wafer. Using EBL, large windows are defined and then aligned with micrometer precision to the pattern on the device layer side. The final etching uses RIE for the hard mask, HCl:H$_3$PO$_4$ for the InP substrate, BHF for the SiN, and H$_2$O:H$_2$SO$_4$:H$_2$O$_2$ for the InGaAs. Lastly, the sample is dried using critical point drying. The PhC has a hexagonal lattice with a lattice constant a = 765 nm and a hole diameter d = 184 nm. The diameter of the PhC is approximately 120 $\mu$m and therefore fully covers the core of a 105 $\mu$m multimode fiber. Indents are placed at the position of the outer diameter of the fiber (125 $\mu$m) which act as breaking points for the transfer to the fiber.

To transfer the optical structures to the fiber tip, the wafer is mounted on a holder, which allows access to the PhC structures from both sides with the fiber end-face. The holder is placed under a microscope camera. The cleaved multi-mode fiber (105 $\mu$m core, 0.22 NA) is mounted vertically underneath the PhC structure on a movable stage and is connected to a white light source to locate the fiber core with the camera. The fiber is moved up through the large hole in the substrate, and the core is aligned with the center of the PhC structure. While approaching, the distance and the optical response of the PhC structure can be evaluated by measuring the reflection spectra with an Optical Spectrum Analyzer (OSA). To finish the transfer, the fiber facet is brought in contact with the suspended structure to break the four lateral supports.

\subsection{Photonic crystal biosensor}
The SiN photonic crystals used for biosensing are fabricated on a 500 $\mu$m thick fused silica substrate. The process begins with the deposition of 410 nm SiN by PECVD at 300 $^\circ$C. This is followed by spin coating of PMMA A4 950K as a resist, and a conductive polymer AR-PC 5090.02 (Electra 92), to avoid charging during the EBL step. Then, an array of markers is defined using EBL, electron beam evaporation of 2 nm/50 nm Ti/Au, and lift-off in acetone. For the second EBL step, a ZEP520A resist is used followed by Electra 92. The array of gold markers is used by the EBL to measure the focus distance to make a height map for the subsequent PhC patterning. After the removal of Electra 92 and the development of the resist, the pattern is transferred into the SiN by RIE with a CHF$_3$ chemistry until a depth of 200 nm is reached. The resist is then removed by an O$_2$ plasma. Finally, the sample is diced into 3 $\times$ 4 mm$^2$ rectangles. 

The PhC has a hexagonal lattice with a lattice constant a = 1140 nm and a hole diameter d = 570 nm. The area of the PhC is a 1 mm diameter circle, which is slightly larger than the light spot in the experimental setup. To expose the PhC structure to different liquids, each die is glued face-up onto a microscope glass slide and is cleaned with air plasma before a flow cell of matching size is glued on the microscope slide (Ibidi sticky-Slide I Luer with 0.8 mm channel height). 

\section{Experimental setup}
The experimental setup used for the experiments is mostly the same, with a few exceptions outlined below. Initially, a coupler is used to direct light from the light source (SLS201L, Thorlabs) onto the sensor and collect its reflectance. The sensor response is then guided by a second coupler to an Optical Spectrum Analyzer (OSA) with a resolution set at 1 nm, as well as to the multispectral readout unit. To launch the light from the coupler fiber port to the multispectral chip, a pair of lenses is employed. The first lens (f=25 mm) collimates the light from the fiber, while the second lens (f=75 mm) focuses the light onto the multispectral chip. This configuration results in a threefold magnification from the fiber end-face to the readout chip.

In the temperature and biosensing experiments, fibers and compatible couplers with a core size of 400 $\mu$m and a numerical aperture (NA) of 0.39 are employed. For the fiber-tip refractive index sensing, fibers and couplers with a core size of 105 $\mu$m and an NA of 0.22 are used. In both temperature and refractive index sensing, the couplers split the light in a 50:50 ratio. However, for the biosensing experiment, a 90:10 coupler is used as the second coupler to maximize the power launched into the multispectral readout chip. The lamp power coupled to the fiber was 10 mW for the temperature and biosensing experiment and 0.4 mW for the fiber-tip refractive index experiment. 

In the biosensing experiment, where the sensing element is not in contact with the fiber optic, the illumination of the PhC structure in the flow cell is achieved from the bottom. To accomplish this, the light from the fiber is collimated using a lens (f=25 mm) and passes through a 1480 nm long-pass filter. Subsequently, the light is focused onto the PhC structure by a second lens (f=60 mm), after which it passes through the microscope slide and the glass substrate.

\begin{figure}[htbp]
\centering
\includegraphics[width=0.99\linewidth]{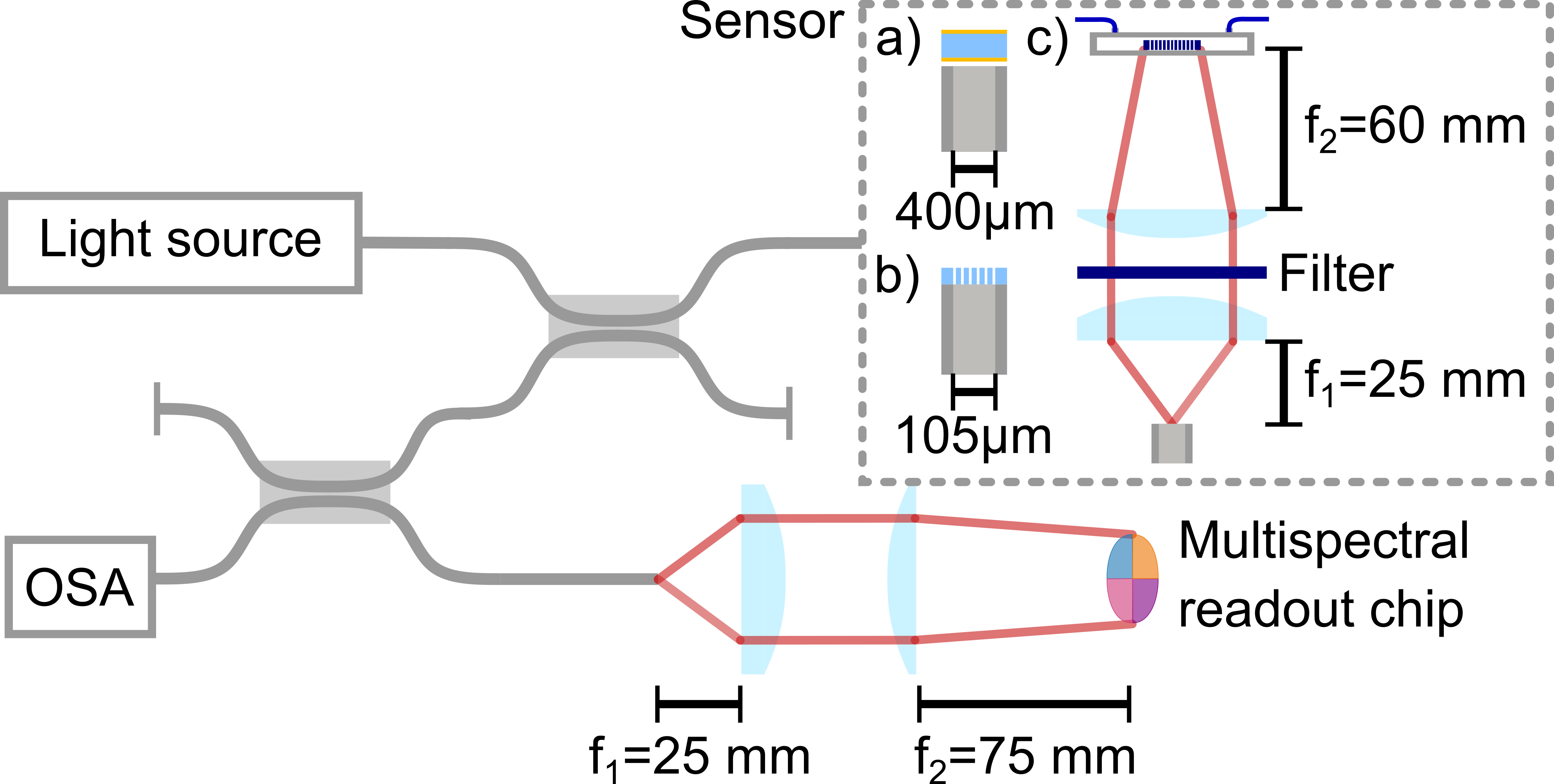}
\caption{Experimental setup: Light is coupled into sensors (a) FP for temperature, (b) Fiber PhC for concentration, and (c) PhC structure for biosensing. The reflection is directed into the OSA and the multispectral readout unit. The parameter of interest is estimated by retrieving changes in the photocurrents produced by the alteration in the reflection spectrum of the fiber optic sensor.}
\label{figS1:setup}
\end{figure}

\section{Estimation model}
For parameter estimation based on data collected with the multispectral sensor readout, we employ an in-house MATLAB\textsuperscript{\textregistered} code. The estimation algorithm utilizes a calibration process to build curves of photocurrents for some values of the parameter to be estimated, as shown in Figure 1(c) of the primary manuscript. These calibration curves represent an N-dimensional parametric equation as a function of the measurand of interest, $I(\lambda_S) = {I_1 (\lambda_S),I_2 (\lambda_S),I_3 (\lambda_S),I_4 (\lambda_S)}$. By knowing the function $I(\lambda_S)$ and the measured photocurrents from each spectral channel ($M = {M_1,M_2,M_3,M_4}$) at a specific peak wavelength, the value of the parameter can be retrieved the value by solving $I(\lambda_S)=M$. To estimate the wavelength of the peak, which depends on the sensor measurand, a point on $I(\lambda_S)$ that has the smallest distance ($D(\lambda_S)$) to M has to be found. The finding of the point is done iteratively using Newton’s method to minimize $D(\lambda_S)$.

\section{Experimental protocols}
\subsection{Temperature sensing}
For the temperature sensing experiment, the FP sensor and a thermistor (GA10K

3MCD, TE connectivity) are positioned within a custom-made aluminum block, with the thermistor serving as a temperature reference sensor. The block is designed to minimize environmental temperature fluctuations, ensuring a stable measurement environment. Precise temperature control is achieved by mounting the block on a Peltier element. During the experiment, the initial temperature is maintained at 25 $^\circ$C for one hour. Subsequently, the temperature is increased by 5 $^\circ$C every hour, reaching 65 $^\circ$C after 9 hours. Following this, the temperature is reduced to 60 $^\circ$C and held stable for 1 hour. The temperature is then decreased similarly three more times until it reaches 30 $^\circ$C. The peak wavelength of the OSA spectra, the temperature readings from the thermistor, and the photocounts of each detector are continuously recorded throughout the entire experiment. The obtained data is used for calibration and evaluation of the multispectral readout unit and estimation model. The calibration data consists of the first nine plateaus representing the ascending temperature phase, while the remaining plateaus, including the transient periods, are reserved for testing the performance of the system. A detailed visualization of the time trace can be found in Section \ref{subsec:timetrace} of this document, in Fig. \ref{figS2:TSensing}. 

\subsection{Fiber-tip refractive index sensing}
To expose the fiber-tip sensor to solutions with different refractive indices, the sensor is immersed in mixtures of isopropanol (IPA) in distilled water (DI) at constant temperatures. For calibration of the estimation model, the sensor is immersed in solutions with concentrations of 100, 80, 60, 40, 20, and 0 $\%_{(m/m)}$. Similarly, for the test data, the sensor is immersed in solutions with concentrations of 10, 30, 50, 70, and 90 $\%_{(m/m)}$. The sensor is immersed in each solution for one minute, and then cleaned with pure IPA before being immersed in the next solution. During this time, the peak wavelength of the spectra measured by the OSA and the photocounts of each detector in the multispectral sensor are recorded. The refractive index of each solution at 1550 nm is calculated using the Gladstone-Dale (GD) equation \cite{Saunders2016-ss}.

\subsection{Biosensing} \label{subs:bioprotocol}
The flow cell, containing the PhC structure, allows for the introduction of various liquids by pumping them into the cell at a flow rate of 0.5 mL/min using a syringe pump. During the bioassay, the surface of the PhC structure is functionalized by physisorption with immunoglobulin G (IgG) molecules as part of the biosensing experiment shown in Fig. 5(d) of the primary manuscript. For the functionalization with IgG, the biosensor is exposed to a 300 nM IgG (from rabbit serum) in PBS solution for 50 minutes. The adsorption of the receptor molecules on the SiN surface is evident by the observable change in the peak wavelength, which is non-reversible upon flushing with PBS solution. As a blocking step, the biosensor is exposed to a casein solution (Casein Blocking Buffer 10x from Sigma-Aldrich, 100x diluted). This ensures that the available sites of the SiN surface are filled with the casein molecules, thereby avoiding unspecific binding of other biomolecules on the surface. The control experiment shown in the Section \ref{subs:control} of this document proves that the achieved surface functionalization is specific to the target molecule anti-IgG, as the repeated exposure to the IgG solution does not show a red-shift and therefore no unspecific binding of the IgG molecules after functionalization (see Figure \ref{figS5:Control_B}). During the biosensing experiment, the biosensor is exposed to the analyte solution with anti-IgG at 100 nM concentration (from goat antibody to rabbit-IgG) in PBS solution. Finally, to prove that the wavelength shift is non-reversible, a rinsing step with a PBS solution is performed.

\section{Sensitivity and Cramér-Rao lower bound}
\subsection{Approximate expressions for the Cramér-Rao lower bound}
The CRLB on the wavelength estimation (Eq. 1 in the primary manuscript) depends on the derivatives of the photocurrents: $\frac{\partial \overline{I}_i}{\partial \lambda_S}=\int_{}^{} R_i (\lambda)\frac{\partial R_S}{\partial \lambda_S} P_\lambda(\lambda)d\lambda$. To provide simple estimates for these, we approximate the responsivity as $R_i(\lambda)\simeq\frac{e}{hc}\lambda_S A(\lambda)$, where $A(\lambda)$ is the absorptance of the detector (including the filtering structure), $h$ the Planck's constant, and $e$ is the electron charge. We assume that the power spectral density is constant, $P_\lambda(\lambda)=P_\lambda$. We also assume that $R_S(\lambda;\lambda_S)$ has a spectral feature (e.g. a peak or dip) centered at $\lambda_S$ and that it translates along the wavelength axis according to $\lambda_S$. We can then write $\left| \frac{\partial \overline{I}_i}{\partial \lambda_S} \right|=\frac{e}{hC}\lambda_SP_\lambda\left| \int_{}^{} A_i(\lambda)\frac{\partial R_S}{\partial \lambda_S} d\lambda\right|=\frac{e}{hC}\lambda_SP_\lambda\left| S_i(\lambda_S) \right|$, where $S_i(\lambda_S)\equiv \int_{}^{}A_i(\lambda)\frac{\partial R_S(\lambda;\lambda_S)}{\partial \lambda_S}d\lambda=-\int_{}^{}A_i(\lambda)\frac{\partial R_S(\lambda;\lambda_S)}{\partial \lambda}d\lambda$ is an adimensional sensitivity. In the ideal case of no optical loss, assuming that $A_i$ and $R_S$ consist of single Lorentzian lines centered at $\lambda_i$ and $\lambda_S$, with the same full-width half-maximum $\Delta\lambda_{FWHM}$, it is easy to show that $S_i(\lambda_S)$  is zero for $\lambda_i=\lambda_S$ and takes a maximum value of $\sim$0.5 (independent of $\Delta\lambda_{FWHM}$) for $\lambda_i\simeq\lambda_S\pm\Delta\lambda_{FWHM}/2$, i.e. when the readout channels are positioned on either side of the sensor resonance. Reducing or increasing the width of the readout resonances reduces the peak value of $S$. This shows that the highest sensitivity and lowest wavelength imprecisions are obtained for matched sensor/readout linewidths. Generally, as $A_i$ and $R_S$ are bound between 0 and 1, each spectral band, where $A_i$ and $R_S$ both show a peak, can contribute at most 1 to the sensitivity, independent of their widths.

In the simplest case of single-channel readout, the Cramér-Rao bound is given by $\sigma\mathrm{}_{\lambda_S}^{CR}=\frac{\sigma_I}{\left| \frac{\partial \overline{I}}{\partial \lambda_S} \right|}=\frac{\sigma_I}{\frac{e}{hC}\lambda_SP_\lambda\left| S_i(\lambda_S) \right|}=\frac{P_{min}}{P_\lambda\left| S_i(\lambda_S) \right|}$, where $P_min$ is the detector's noise-equivalent power. For a single, optimally detuned readout channel, ($\left| S_i(\lambda_S) \right|\sim$ 0.5) and $\sigma\mathrm{}_{\lambda_S}^{CR}=2 P_min/P_\lambda$, which has a simple physical interpretation: The minimum wavelength shift $\delta\lambda_{min}$ that can be measured is such that the power of the source in that wavelength interval is of the order of the minimum power the detector can measure, $P_\lambda \delta\lambda_{min}\sim P_{min}$ (the additional factor 2 is due to the use of a single sideband). Any loss in the optical system and readout (including the loss due to the spatial multiplexing in the detector array and any detector inefficiency) can be incorporated into $P_\lambda$, which then becomes the power spectral density incident on the detector. By placing a second detector at the opposite side of the sensor resonance, the CRLB is slightly improved to $\sigma\mathrm{}_{\lambda_S}^{CR}=\frac{\sigma_I}{\sqrt{2} \left| \frac{\partial \overline{I}_i}{\partial \lambda_s} \right|}$. However, as seen from Eq. 1 of the primary manuscript, splitting up the readout in more channels with smaller linewidth (i.e. moving towards a spectrometer-based read-out) tends to degrade the wavelength imprecision, as the photocurrent scales with the linewidth. In practice, additional optical loss is usually associated with narrower linewidth (e.g. due to the need of reducing the étendue), which makes the scaling even worse.

\subsection{Case of shot noise}
The considerations of the primary manuscript apply to the case where thermal noise in the readout circuit is dominant, which is the most likely situation in a practical application setting. For dominating shot noise (in the limit of large photon numbers), Eq. 1 of the primary manuscript is modified to

\begin{equation}
\sigma\mathrm{}_{\lambda_S}^{CR}=\frac{1}{\sqrt{\sum_{i=1}^{N}\frac{1}{\sigma\mathrm{}_{I_i}^{2}}\left( \frac{\partial \overline{I}_i}{\partial \lambda_S} \right)^2}}=\frac{\sqrt{2e\Delta f}}{\sqrt{\sum_{i=1}^{N}\frac{1}{\overline{I}_i}\left( \frac{\partial \overline{I}_i}{\partial \lambda_S} \right)^2}}
\label{eq:shot}
\end{equation}

(where $\Delta f$ is the measurement bandwidth), from which we see that, in an ideal optical system, the wavelength imprecision is independent of the number of readout channels. However, in both dispersive spectrometers and multispectral arrays, improving the spectral resolution requires reducing the étendue, so that using more spectral channels with improved resolution leads to a degradation of the imprecision. 

\subsection{Generalization of the Cramér-Rao lower bound to more complex spectral changes}
We have so far referred to the estimation of the wavelength $\lambda_S$, assuming that the parameter of interest $x$ is mapped into a spectral shift of a single resonance, as in most existing resonant sensors. However, in a more general statement of the sensing problem, the lower bound on the imprecision in the parameter $x$ is given (for thermal noise) by

\begin{equation}
\sigma\mathrm{}_{\lambda_S}^{CR}=\frac{1}{\sqrt{\sum_{i=1}^{N}\left( \frac{\partial \overline{I}_i}{\partial x} \right)^2}}
\label{eq:general}
\end{equation}

where $\frac{\partial \overline{I}_i}{\partial x}=\int R_i (\lambda) \frac{\partial R_S}{\partial x} P_\lambda (\lambda) d \lambda$  can describe more general changes in the spectrum (e.g. increased/decreased reflection in certain bands). This allows benchmarking the performance of different sensing systems, including those where the transducer and the readout unit have more complex spectral lineshapes. 

This approach based on the CRLB is very powerful as it allows comparing the ultimate performance of a variety of optical sensing systems, independently of the fitting/data analysis methods used to build prediction models. 

\section{Temperature sensing experiments} 

\subsection{Temperature sensing time trace} \label{subsec:timetrace}
\begin{figure}[h!]
\centering
\includegraphics[width=0.9\linewidth]{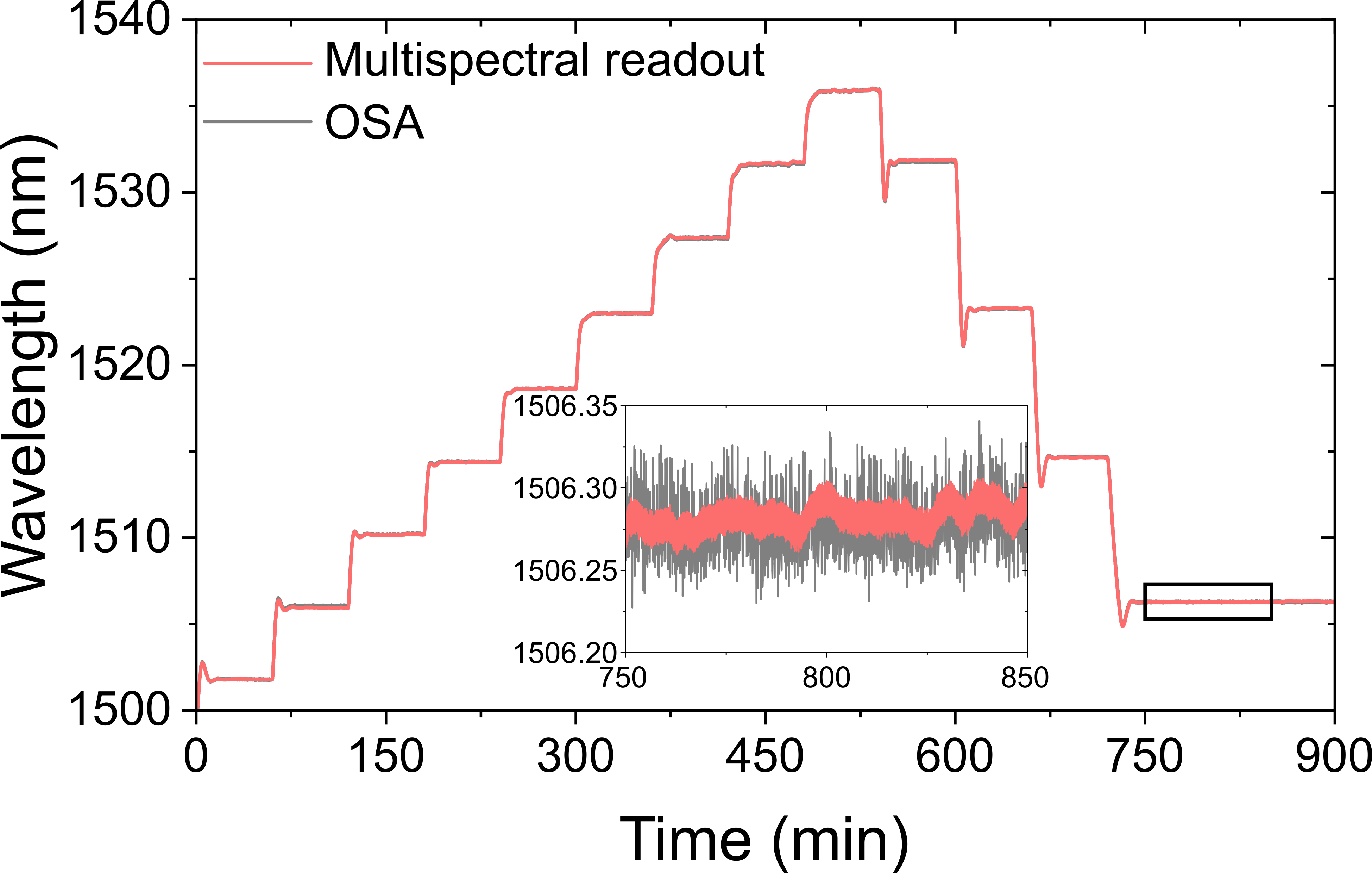}
\caption{Time trace of the resonance wavelength obtained from the OSA and multispectral readout during a temperature experiment. The inset provides a zoomed-in view from 750 to 850 minutes to highlight the imprecision of both approaches.}
\label{figS2:TSensing}
\end{figure}

In Fig. \ref{figS2:TSensing}, the time trace of the wavelength, calculated using the OSA, and estimated with the multispectral readout, is presented. The temperature was systematically increased from 25 °C to 65 °C and then decreased from 60 °C to 30 °C, following the method described in the Methods section. The plot illustrates that the wavelength estimated by the multispectral readout closely matches the OSA measurements, not only during the plateaus but also during the transient periods. Furthermore, the inset demonstrates that the multispectral readout yields more accurate results.

\subsection{Cramér-Rao lower bound in the experimental setting}
\begin{figure}[h!]
\centering
\includegraphics[width=0.9\linewidth]{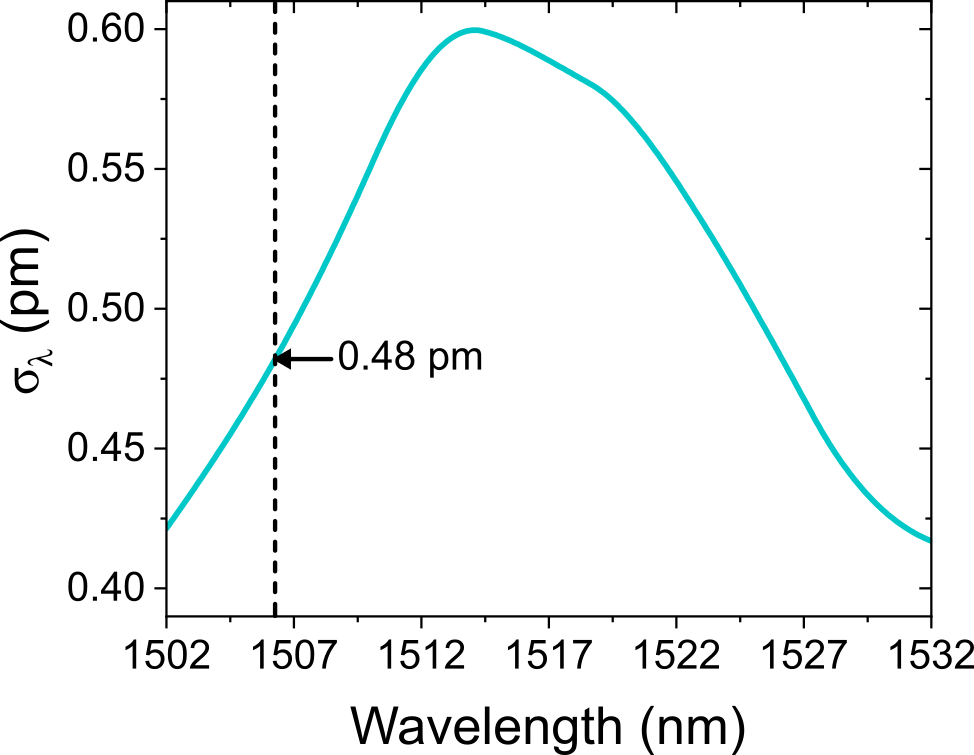}
\caption{CRLB calculated from the sensitivities measured in the temperature sensing experiment and the experimental SNR corresponding to an integration time of 4.5 s.}
\label{figS3:CRL_T}
\end{figure}

The CRLB for wavelength estimation shown in Fig. 1(d) of the primary manuscript refers to the case of Lorentzian lineshapes for the readout channels and an ideal Fabry-Perot cavity. To compare it to the experimental wavelength imprecision, we calculate the CRLB (Eq. 1 of the primary manuscript) with the sensitivities measured in the temperature sensing experiment (slopes of the counts vs wavelength data in Fig. 3(d) of the primary manuscript), and the experimental SNR corresponding to an integration time of 4.5 s. The resulting $\sigma\mathrm{}_{\lambda_S}^{CR}$ (Fig. \ref{figS3:CRL_T}) is a function of wavelength, as the sensitivities vary with wavelength. At the temperature and wavelength (T=30 $^\circ C$, $\lambda$=1506.3 nm), for which the Allan deviation of Fig. 3(d) was measured, we obtain $\sigma\mathrm{}_{\lambda_S}^{CR}$=0.48 pm, which is very close to the experimental wavelength imprecision of 0.6 pm (minimum of Allan deviation of Fig. 3(d)). This proves that the estimation algorithm works optimally also with experimental data and provides a wavelength estimation with imprecision close to the fundamental CRLB limit.

\subsection{Use of a light-emitting diode as a light source}
\begin{figure}[h!]
\centering
\includegraphics[width=0.9\linewidth]{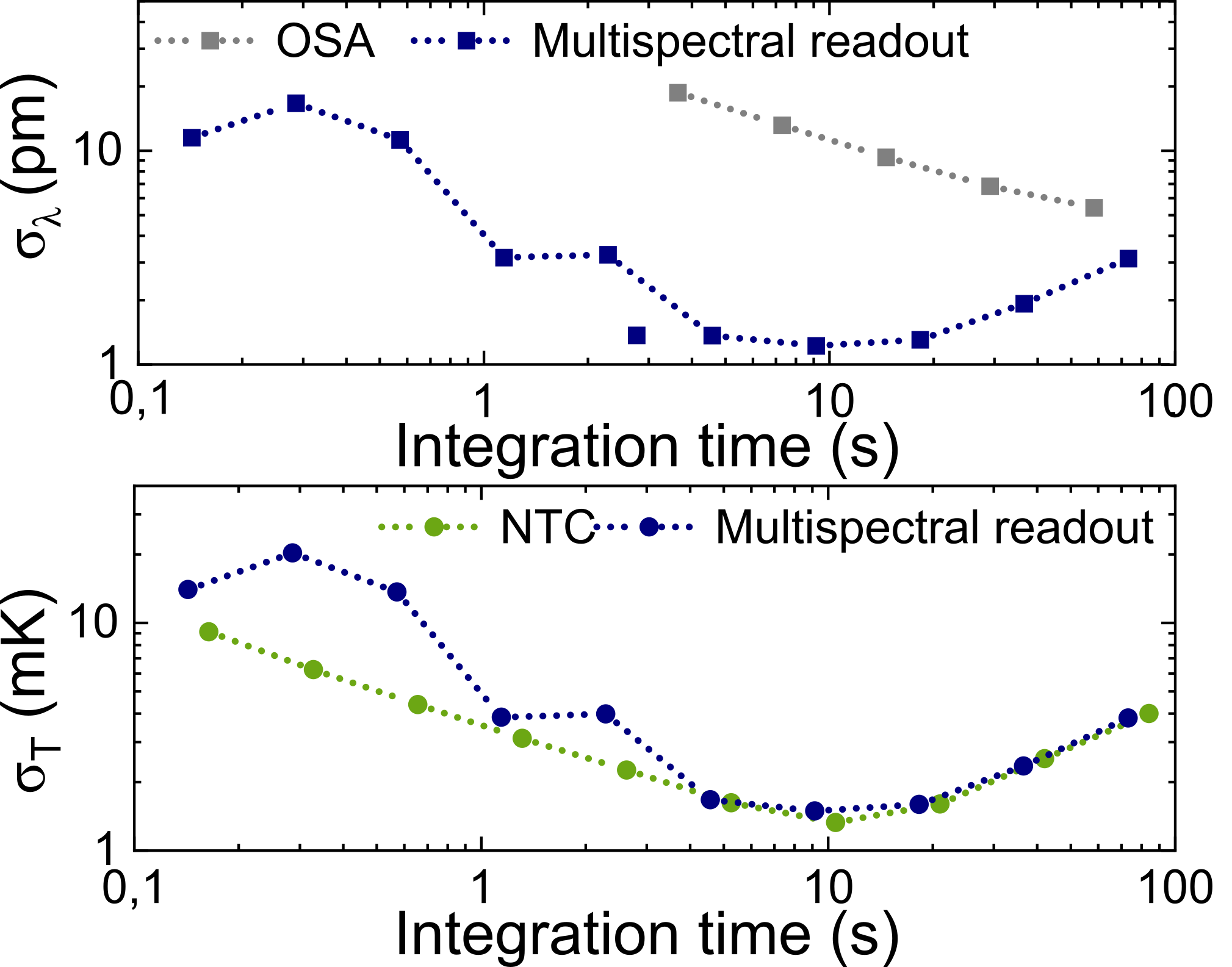}
\caption{Allan deviation analysis of wavelength and temperature for multispectral readout, Optical Spectrum Analyzer (OSA), and reference temperature sensor (NTC) when a light-emitting diode (LED) is used as the light source.}
\label{figS4:LED_T}
\end{figure}

To evaluate the feasibility of using a different light source for the proposed approach, a light-emitting diode (LED) was used instead of a halogen lamp in similar experiments to those described in the primary manuscript. Figure \ref{figS4:LED_T} illustrates the Allan deviation obtained when a Fabry-Pérot temperature sensor is used in combination with an LED ($\lambda_C$  = 1550 nm, FHWM = 102 nm, P = 138 $\mu$W coupled to the fiber). The plot demonstrates that using the LED as the light source allows for a minimum wavelength imprecision of 1.4 pm and a minimum temperature imprecision of 1.7 mK. Although these values are slightly higher than those obtained with the halogen lamp, they are still comparable to the imprecision of the reference thermistor.

\section{Biosensing experiments}
\subsection{Control experiment} \label{subs:control}
\begin{figure}[h!]
\centering
\includegraphics[width=0.9\linewidth]{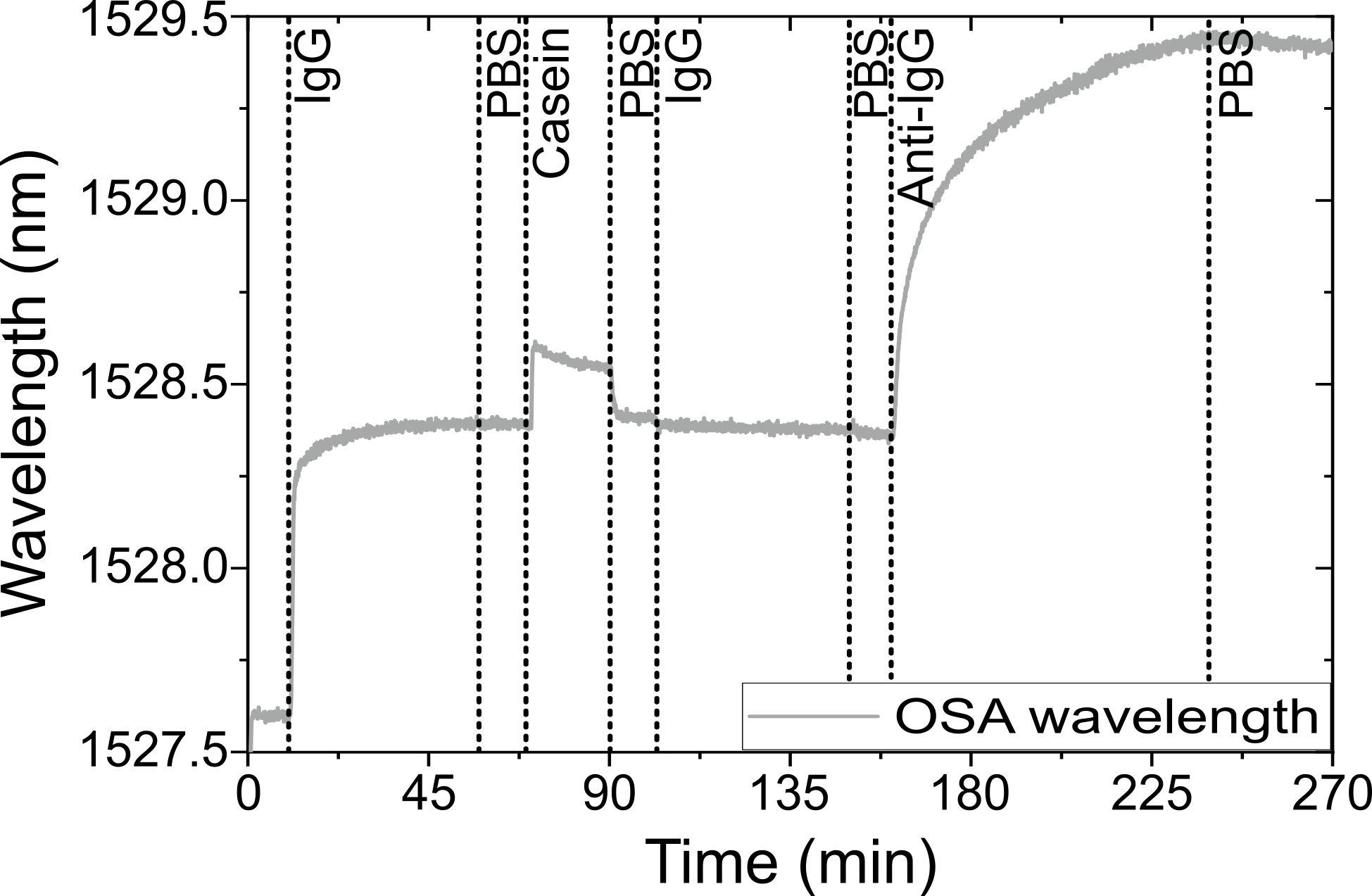}
\caption{Time trace of the resonance wavelength obtained from the OSA during a biosensing experiment, where the functionalized sensor is exposed to IgG and subsequently to Anti-IgG to assess the specificity of the functionalization.}
\label{figS5:Control_B}
\end{figure}

To prove the specificity of the functionalization of the biosensor, a control experiment was carried out on a nominally identical copy of the PhC device. The results are depicted in Fig. \ref{figS5:Control_B}. The surface was functionalized in the same way (see Section \ref{subs:bioprotocol} of this document) with an IgG in PBS solution by physisorption (300 nM, 15 – 60 min) and a blocking step with a casein solution (75 – 95 minutes). After rinsing with PBS (until 100 min), the biosensor surface is again exposed to the same IgG in PBS solution (100 – 150 min). From the measured OSA spectrum, it is evident that there is no wavelength shift due to the exposure to biomolecules as to be expected for unspecific interactions. Consequently, the interaction with the target molecule anti-IgG in the next step of the assay (100 – 240 min) shows a clear response that corresponds to a specific interaction.

\bibliographystyle{unsrt}
\bibliography{sample}
